\documentclass[%
reprint,
amsmath,amssymb,
aps,
]{revtex4-2}

\usepackage{graphicx}
\usepackage{dcolumn}
\usepackage{bm}

\begin{document}
	
	
	\title{Amplification of light scattering  in arrays of  nanoholes by  plasmonic absorption-induced transparency}

	\author{Sergio G. Rodrigo}%
	\email{sergut@unizar.es}
	\affiliation{%
		Centro Universitario de la Defensa,\\
		Ctra. de Huesca s/n, E-50090 Zaragoza, Spain
	}%
	\affiliation{%
	Instituto de Nanociencia y Materiales de Aragón (INMA),CSIC-Universidad de Zaragoza,\\
	50009 Zaragoza, Spain
}%

	\begin{abstract}
		 Absorption induced transparency is an optical phenomenon that occurs in metallic arrays of nanoholes when materials featuring narrow lines in their absorption spectra are deposited on top of it. First reported in the visible range, using dye lasers as cover materials, it has been described as transmission peaks unexpectedly close to the absorption energies of the dye laser. In this work, amplification of light  is demonstrated in the active regime of absorption induced transparency. Amplification of stimulated emission can be achieved when the dye laser behaves as a gain material. Intense illumination can modify the dielectric constant of the gain material,  which in turn, changes the propagation properties of the plasmonic modes excited in the hole arrays, providing both less damping to light and further feedback, enhancing the stimulated emission process.
	\end{abstract}
\maketitle
	
\section{Introduction}
Lasers are ubiquitous devices in technology with strong presence in our everyday life~\cite{GarmireOptExp13}. The idea of down-scaling lasers to the nanoworld sparked the imagination of scientists since the beginning of lasers. Nanotechnology cleared the way to this goal, in part supported by a plenty of discoveries in the field of plasmonics. Plasmonics exploits a great variety of electromagnetic (EM) modes, bound to the nanostructures in subwavelength volumes.  In metals the strong confinement of light is due to Surface Plasmon Polaritons (SPP). The phenomena in which surface plasmons are involved are many and of a very diverse nature. For instance, in the quest of nanolasers metals and colloidal solutions of quantum dots have been combined to recently demonstrate plasmonic distributed-feedback lasers~\cite{NorrisACSNano21} and, dye molecules, metals and silica-based photonic crystals have been used to create a 3D plasmonic laser~\cite{StockmanOptExp21}. 

Plasmonic lasers requires not just of a gain material to work but also of platforms that provide the necessary feedback to build the lasing action. Lattices of metallic nanoobjets (e.g. nanoparticles deposited on a substrate or holes drilled on metal films) are the basic blocks for laser operation at the nanoscale. These are able to concentrate light in very hot spots distributed across the whole lattice size forming regular patterns. These kind of systems can be used to generate amplification and lasing, among others~\cite{ZhouNatNano13,WuestnerPRL10,MaraniNJP12}. 

Regular arrangement on nanoobjects can  give raise to emerging optical properties of the whole system that are not present in the constituent materials. These kind of structures are so-called metasurfaces and their optical response are described through effective theories, where the nanoobjects play the role of "atoms". Metasurfaces are interesting for many applications in extreme regimes of light-matter interaction, where different forms of non-linear response are present~\cite{Hess2012}. 

We investigate the gain response of a metasurface which collective behavior is due to the  Absorption Induced Transparency (AIT) effect. AIT was discovered in arrays of nanoholes and was described as unexpected peaks in their transmission spectrum, after the incorporation of a dye laser on the metal surface~\cite{HutchisonAnge11} . An opaque film (the metallic array of holes) becomes translucent after incorporation of the solution containing the dye laser. The physical mechanism was initially puzzle, as AIT occurr at frequencies of high absorption because the presence of the molecules.

Two physical mechanisms explain AIT. The first  was reported in Ref.~\cite{RodrigoPRB13} and explained AIT as due to a strong modification of the propagation constant of light inside the holes.  The second mechanism is related with the excitation of SPPs, ultra-confined at the layer where the molecules are located~\cite{ZhongACSNano16}. Both contribute to AIT and are related with other processes of light scattering in corrugated metal films, and in particular with the phenomenom of Extraordinary Optical Transmission (EOT)~\cite{RodrigoIEEE16}.

The AIT phenomenon was observed in the visible range for the first time, although later on it was predicted to occur outside that frequency window~\cite{RodrigoPRB13}. The theoretical prediction was ultimately verified in the terahertz regime, where the role of the electronic transitions of dye molecules in the visible was played by phonon resonances in LiF~\cite{AcostaAdvOptMat17}.

Apart from a few studies focused on fundamental aspects of light-matter interactions~\cite{PetoukhoffNatComm15} or on the detection of gases in the THz regime~\cite{RodrigoEPJ16}, AIT has been elusive in many experiments. Let us say that only a few experiments in strong coupling found effects ascribed to AIT~\cite{ZhongJPC20}, while many others did not do, although they were conducted for similar conditions of filling/covering~\cite{NorrisNanoLett19,DeLeonNatPhot20}.

In this work we theoretically demonstrate the amplification of stimulated light relying on active AIT. The rest of the work is organized as follows: the basic elements of the theoretical framework is discussed in Section~\ref{materials}, the main results devoted to AIT in the active regime are provided in Section~\ref{results} and, finally, we will end with the conclusions in Section~\ref{conclusions}.

\begin{figure*}
	\centering\includegraphics[width=2.0\columnwidth]{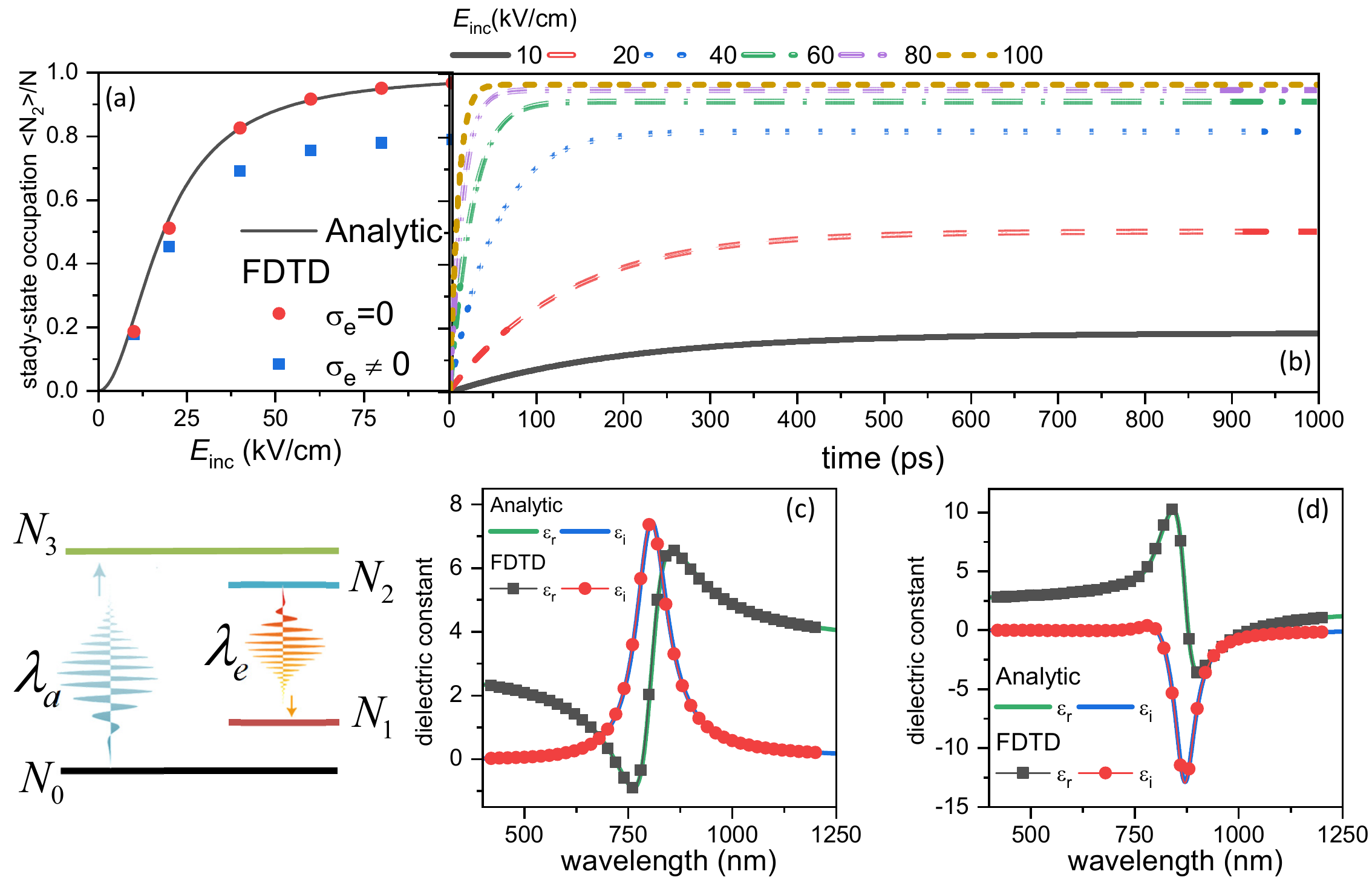} \caption{The optical response of the infrared dye laser IR-140 used in this work as gain material can be approximated as a four-level atomic system, shown in the schematics (for further details, see main text ). The simulated material consist of a polymeric matrix of PMMA where the IR-140 dye is introduced. The host medium has refractive index $n_h =$1.62. (a) Comparison between numerical simulations (symbols) and the analytic result (solid line) of the normalized steady-state occupation density of the upper emission state ($\langle N_2 \rangle/N$) in a 4~nm thin slab of gain material when continuously pumped by plane wave at $\lambda_a=805$~nm, as a function of the electric field amplitude of the incident field. Square symbols are full calculations using Maxwell-Bloch equations, while circular symbols show and approximation where null overlap between emission and absorption bands is assumed. (b) For this approximation, time evolution of $\langle N_2 \rangle/N$ for different pump intensities. (c)-(d) Steady-state dielectric constant ($\varepsilon_r + \imath \varepsilon_i$) inside the 4~nm thin slab of gain material for: (c) weak pumping and (d) intense illumination, under an incident electric field of $E_{inc}=40$~kV/cm, which produces a steady-state population of the emission state $\langle N_2 \rangle/N=0.83$, for $\sigma_e=0$.} \label{fig0}
\end{figure*}

\section{Materials and Methods}\label{materials}
The dye laser IR-140 has been chosen as the gain material to illustrate the principles of active AIT.  This dye molecule has demonstrated to perform better than other dye lasers in terms of stability and photobleaching, at the expense of working in the infrared~\cite{ZhouNatNano13}. Note however that our results can be applied to other gain materials, like the promising quantum dot based nanoplatelets~\cite{MarianneChemMat19}.   

The optical properties (including gain) of the IR-140 dye laser can be approximately described using a four-level atomic model~\cite{SperberOptQuantum88}. Schematically shown in Fig.~\ref{fig0}, IR-140 optical response is fully characterized by the following material parameters: host refractive index $n_{h}=1.62$ (actually, the model simulates the infrared IR-140 dye laser within a polymer PMMA matrix); population of the ground state $N_0=0.6$~molecules/nm$^3$; absorption and emission wavelengths $\lambda_a=805$~nm and $\lambda_e=870$~nm; absorption and emission dipole strengths (coupling constants) $\sigma_a=6.898 \times 10^{-8} \space C^2/kg$ and $\sigma_e=5.152 \times 10^{-8} \space C^2/kg$; resonances half-width $\Gamma_a=0.134$~fs$^{-1}$ and $\Gamma_e=0.075$~fs$^{-1}$. Finally, relaxation times are: $\tau_{32}=\tau_{10}=100$~fs and $\tau_{21}=240$~ps.

Our predictions are based on numerical calculations conducted with an own code of the Finite Difference Time Domain (FDTD) method, which provides a self-consistently approach for solving both Maxwell's and molecular-rate equations based on the Maxwell-Bloch formalism. A complete description of the Maxwell-Bloch formalism can be found in Ref.~\cite{wuestnerPhilo11}. Information about the rest of the details of our FDTD implementation can be found elsewhere~\cite{RodrigoTESIS}. The full temporal evolution of the population densities for a four-level system can be expressed as follows:

\begin{eqnarray}\label{fullmodel}
	\frac{\partial N_3}{\partial t} &=& \frac{\vec W_a}{\hbar \omega_a} \vec E_{loc} -\frac{N_3}{\tau_{32}} \\
	\frac{\partial N_2}{\partial t} &=& \frac{N_3}{\tau_{32}}+ \frac{\vec W_e}{\hbar \omega_3} \vec E_{loc} -\frac{N_2}{\tau_{21}}\\
	\frac{\partial N_1}{\partial t} &=& \frac{N_2}{\tau_{21}}- \frac{\vec W_e}{\hbar \omega_e} \vec E_{loc} -\frac{N_1}{\tau_{10}}\\
	\frac{\partial N_0}{\partial t} &=& \frac{N_1}{\tau_{10}}-\frac{\vec W_a}{\hbar \omega_a} \vec E_{loc} 
\end{eqnarray}
, where $N_i=N_i(\vec r, t)$ are the occupations of the different energy levels. The coupling factor, $\vec W_i=	\frac{\partial \vec P_i}{\partial t}+\Gamma_i \vec P_i$, being $\vec P_i(\vec r, t)$ the polarization densities labeled as $i=a,e$ for absorption and emission respectively. The absorption and emission transitions are accessible to the local electric field, $E_{loc} = [(2 + n_h^2)/3]E$, in the Lorentz approximation through the polarizabilities: 

\begin{equation}\label{polarization}
\frac{\partial^2 \vec P_i}{\partial t^2}+2 \Gamma_i \frac{\partial \vec P_i}{\partial t}+\omega_i^2 \vec P_i = - \sigma_i \Delta N_i \vec E_{loc}
\end{equation}
Here $\omega_i=\frac{2 \pi c}{\lambda_i}$, $\Delta N_a=N_3(\vec r, t)-N_0(\vec r, t)$ and $\Delta N_e=N_2(\vec r, t)-N_1(\vec r, t)$.

The dielectric constant derived from the four-level atomic system is described by a sum of Lorentz terms, corresponding to absorption and emission electronic transitions, and it can be analytically expressed as follows: 

\begin{widetext}
\begin{equation}\label{dconstant}
	\varepsilon(\omega)=\varepsilon_h+ \frac{(2 + n_h^2)}{3}\frac{\Delta N_a}{\varepsilon_0}\frac{\sigma_a
	}{(\omega^2-\Omega_a^{2}+\imath 2 \omega\Gamma_a)} 
    +  \frac{(2 + n_h^2)}{3}\frac{\Delta N_e}{\varepsilon_0}\frac{\sigma_e
	}{(\omega^2-\Omega_e^{2}+\imath 2 \omega\Gamma_e)}.
\end{equation}
\end{widetext}
This mathematical expression permits to obtain the local gain once the population levels are known by other means, for instance aid by the FDTD method.

Figure~\ref{fig0}(a) shows the normalized steady-state occupation density of the upper emission state, $\langle N_2 \rangle/N$, inside a $4$~nm thin slab of gain material when continuously pumped by plane wave at $\lambda_a=805$~nm, as a function of the electric field amplitude of the incident field. Square symbols are obtained with the full model given by Eq.~\ref{fullmodel}~-~\ref{polarization}. These equations can be approximated for a continuous wave illumination where the populations of interest are those of the steady-state. By doing so, the steady state occupation density of the upper emission state results in:

\begin{equation}\label{n2}
\langle N_2 \rangle/N = \frac{1}{1+3\tau/\tau_{21}+\vert E_{sat} /E_{loc}   \vert^2}
\end{equation}
where $E_{sat}^2=\frac{4 \hbar \omega_a \Gamma_a}{\sigma_a \tau_{21}} $. This formula includes two additional approximations: the overlap of the emission band with the absorption frequency is negligible ($\sigma_e=0$) and $\tau=\tau_{32}=\tau_{10}$. 

The analytical result of Eq.~\ref{n2} is shown with a solid line in Fig.~\ref{fig0}(a). To compare with the analytical results another numerical simulation has been conducted.  Circles show the calculations after fully solving the Maxwell-Bloch Eqs.~\ref{fullmodel}~-~\ref{polarization} where the null overlap between emission-absorption band approximation is implemented by setting $\sigma_e=0$.  In the case of IR-140 this approximation does not hold, its emission and absorption bands strongly overlap, as can be inferred analyzing in Fig~\ref{fig0}(a) the different steady-state values for the occupation regarding $\sigma_e$. 

The numerical time-domain simulations conducted with the FDTD method highlight a crucial point. The steady-state occupation of the atomic levels is fundamental to grasp the details of the stimulated process that might eventually produce lasing. However, it is even more important to understand the dynamics of atomic systems, specially when dealing with femptosecond pulses. For that, full time evolution of $\langle N_2 \rangle/N$ is show in Fig~\ref{fig0}(b) for different incident pump intensities. As we can observe, the higher the pump intensity the faster the steady-state is reached. This information will be used later on to determine the time delay between pump and probe laser pulses in pump-probe numerical experiments, which will used to establish the main results of this work.

Finally, Fig.~\ref{fig0} shows the dielectric constant of the $4$~nm thin slab of gain material (c) before and (d) after continuous pumping with an intense laser beam ($E_{inc}= 40$~kV/cm) at the absorption energy of the molecules. The analytical result, Eq.~\ref{dconstant}, is represented by the solid line, where the steady-state population is $\langle N_2 \rangle/N=0.83$ ($\sigma_e=0$). Fig~\ref{fig0}(c) shows the typical profile of a spectral absorption line: the anomalous dispersion wavelength range, where phase velocity would exceed that of the speed of light, is forced by causality to experience high absorption, which explain the peak in $Im(\varepsilon)$. In contrast, anomalous dispersion disappears in Fig~\ref{fig0}(d), being $Im(\varepsilon) < 0$, so optical gain through the suppression of losses is reached within the thin film.

The excellent agreement between numerical and analytical results highlights the validity  of our numerical implementation.  Equation~\ref{dconstant} provides the dielectric response of a four-level system when the atomic populations are known. To obtain the results of Fig~\ref{fig0}(d) FDTD simulations were carried out on the $4$~nm thin slab of gain media.  As explained before, one of the advantages of the FDTD method is that light pulses can be studied, like those generated by femtosecond lasers. In addition, the FDTD method is able to self-consistently treat gain materials in complex plasmonic heterostructures where the local EM field might abruptly change from point to point. 

\section{Results and Discussion}\label{results}
\begin{figure*}
	\centering\includegraphics[width=1.8\columnwidth]{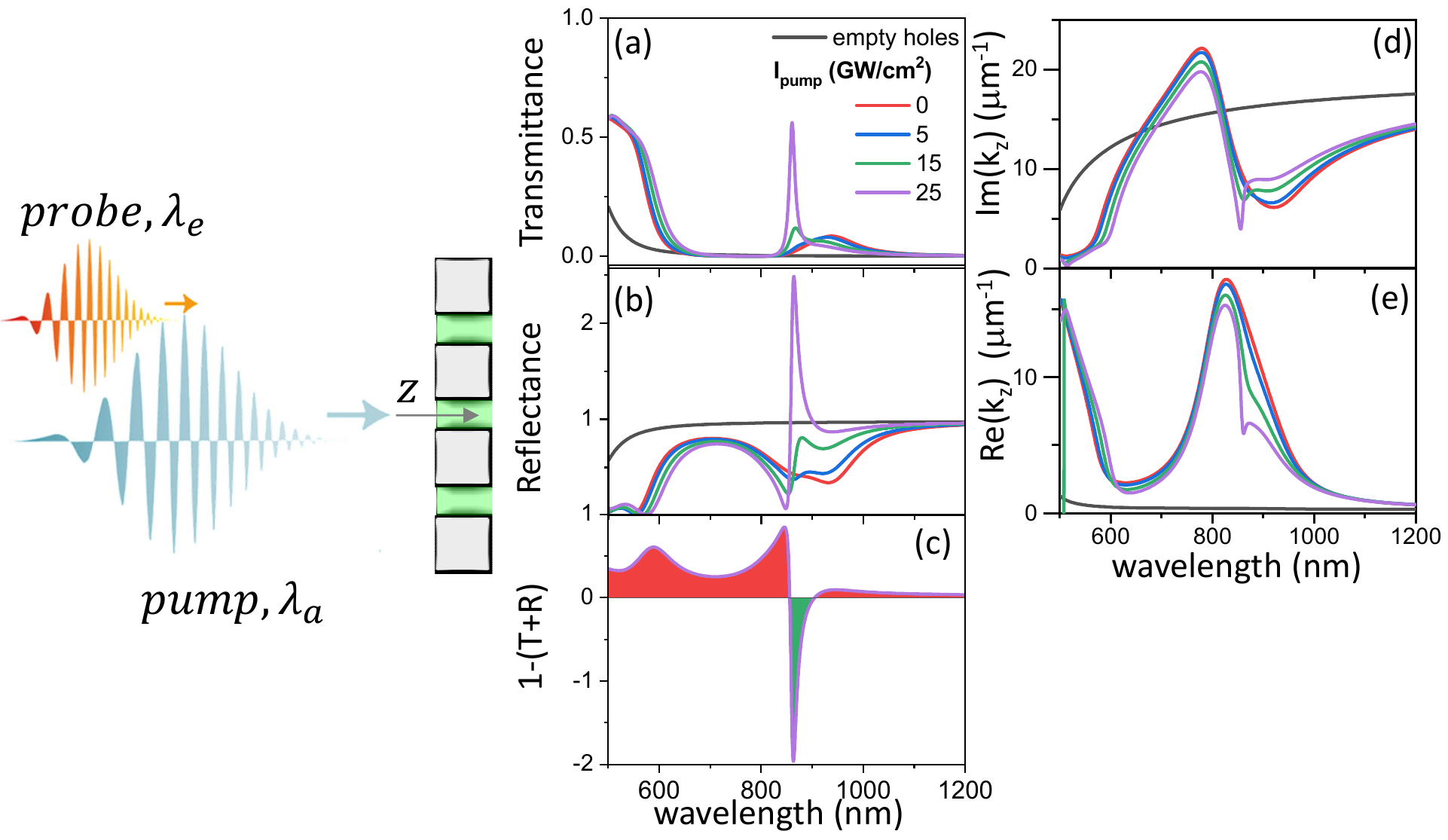} \caption{\textbf{Localized AIT:} (a) Probe transmittance (T) and (b) reflectance (R) as a function of the pump intensity,  for an AIT metasurface of circular holes (diameter, $d = 180$~nm) periodically arranged (period, $p = 250$~nm),  on a silver film (thickness, $h=200$~nm).  The passive case (no IR-140 dye inside the holes) is shown with a solid black line.(c) Light absorbed/emitted by the AIT metasurface calculated as $1-(T+R)$, for the highest pump intensity. (d)-(e) The corresponding imaginary and real part of the propagation constant of holes, $k_z$. } \label{fig1}
\end{figure*}

We investigate the gain response of AIT metasurfaces with the numerical equivalent of a pump-probe spectroscopy experiment, as illustrated in the schematics shown in Fig.~\ref{fig1} and Fig.~\ref{fig2}. The spectral position of the AIT peak, its intensity and spectral width are parameters mainly controlled by the propagation properties of light either inside the holes  and on the surface (as propagating SPPs)~\cite{ZhongACSNano16}. We show that by acting on the characteristic  wavevectors by means of a non-linear process active control of AIT is achieved. 

Pump-probe numerical experiments have been conducted to show the effect of gain in AIT. The metasurface is first pumped with a 150~$fs$ laser beam, central wavelength corresponding to the absorption energy of IR-140 molecules. Finally, a second broadband beam (the probe) is sent after $\Delta t =$800~$fs$, which is weak enough so atomic populations keep unaltered.  On one hand,  the probe hits the system once the gain material has reached the stationary state (see Fig.~\ref{fig0}) and, on the other hand, it is a delay short enough so that the occupation of the emission level seen by the probe is still high.  Note that the scattering coefficients of Fig.~\ref{fig1} and Fig.~\ref{fig2} are that of the probe. The metallic substrate of the AIT metasurface is silver, its optical response taken from experimental tabulated values implemented in our FDTD code by means of a Drude-Lorentz model~\cite{RodrigoPRB08}. 

We will distinguish between localized and surface AIT. In localized AIT the molecules completely fill the holes (Section~\ref{localizedAIT}), whereas a 30~nm thin layer of them covers the metasurface in the case of surface AIT (Section~\ref{surfaceAIT}).

\unskip
\subsection{Localized AIT}\label{localizedAIT}

Figure~\ref{fig1}(a)-(b) shows transmittance and reflectance in a metasurface featuring AIT, which is drilled by a regular array of circular holes (diameter, $d = 180$~nm; period, $p = 250$~nm; and metal thickness, $h=200$~nm). As the intensity of the pump beam increases the population of the upper levels of the molecules increases too, changing the dielectric constant of the dye laser as expected from Eq.~\ref{dconstant}. For that situation, the AIT peak (broadband spectral feature seen in the infrared) evolves towards a narrow emission line, centered at the emission energy of the dye laser, typical in lasing phenomena. Reflection is also affected by the optical gain and a peak shows up at the same wavelength. A better figure of merit about the degree of gain "deposited" by the pump is seen in Fig.~\ref{fig1}(c), which shows $1-(T+R)$.  This expression holds for the absorption in the passive case and it is always positive, i.e., photons are lost from the incident field. In the active situation energy in the matter states (the dye laser) can be released as photons that produces a negative "absorption" by means of stimulated light emission~\cite{siegman86}. The red shaded regions are dominated by absorption.  Into the green shaded region stimulated emission is taking place. This trend is markedly clearer at $\lambda_e$, energy at which photons at the light field substantially exceeds those delivered by the probe.

Enhanced stimulated emission within such a narrow spectral band can be understood by analyzing two mechanisms related to the propagation of light inside the filled holes. For that we calculated $k_z$, which has an analytical expression for circular waveguides~\cite{Jackson75}. The first mechanism is the reduction of the imaginary part of $k_z$ around the emission line of the molecules $\lambda_e$, as shown in Fig.~\ref{fig1}(d), which obviously reduces absorption at that energy. 

The second one, the most intriguing process indeed, is related with the strong modification of the real part of $k_z$. In Fig.~\ref{fig1}(e) an almost ''vertical'' slope develops as the intensity of the pump gets higher and higher values. Therefore a slow down of group velocity of light is produced inside the holes. The diminishing of group velocity has been arged as able to increase stimulated emission in photonic crystals~\cite{SakodaOptExp99a}. We belive that a similar physical response is taken place in our case and might suggest that, like in photonic crystals, low-threshold lasers might be designed based on AIT~\cite{SakodaOptExp99b}. 

Another interesting aspect of active AIT is that in passive media (or low pump powers) AIT occurs in a three-step, non-resonant process~\cite{RodrigoPRB13}. For a resonant process to occur  light must propagate back and forth inside the holes like in classical EOT~\cite{MartinMorenoPRL01}. In passive AIT we demonstrated in Ref.~\cite{RodrigoPRB13} that light first couples with holes, it propagates without being reflected at the second interface, exiting the metasurface as transmitted light out of the holes. Reflection is the light that it is neither absorbed nor transmitted. This trend is broken for high intense illumination in gain materials. The stimulated emission peak at $\lambda_e$ found in Fig.~\ref{fig1}(c) can only be explained if a feedback mechanism makes the photons stay in the system enough duty cicles.  


\unskip
\subsection{Surface AIT}\label{surfaceAIT}
\begin{figure}
	\centering\includegraphics[width=1.0\columnwidth]{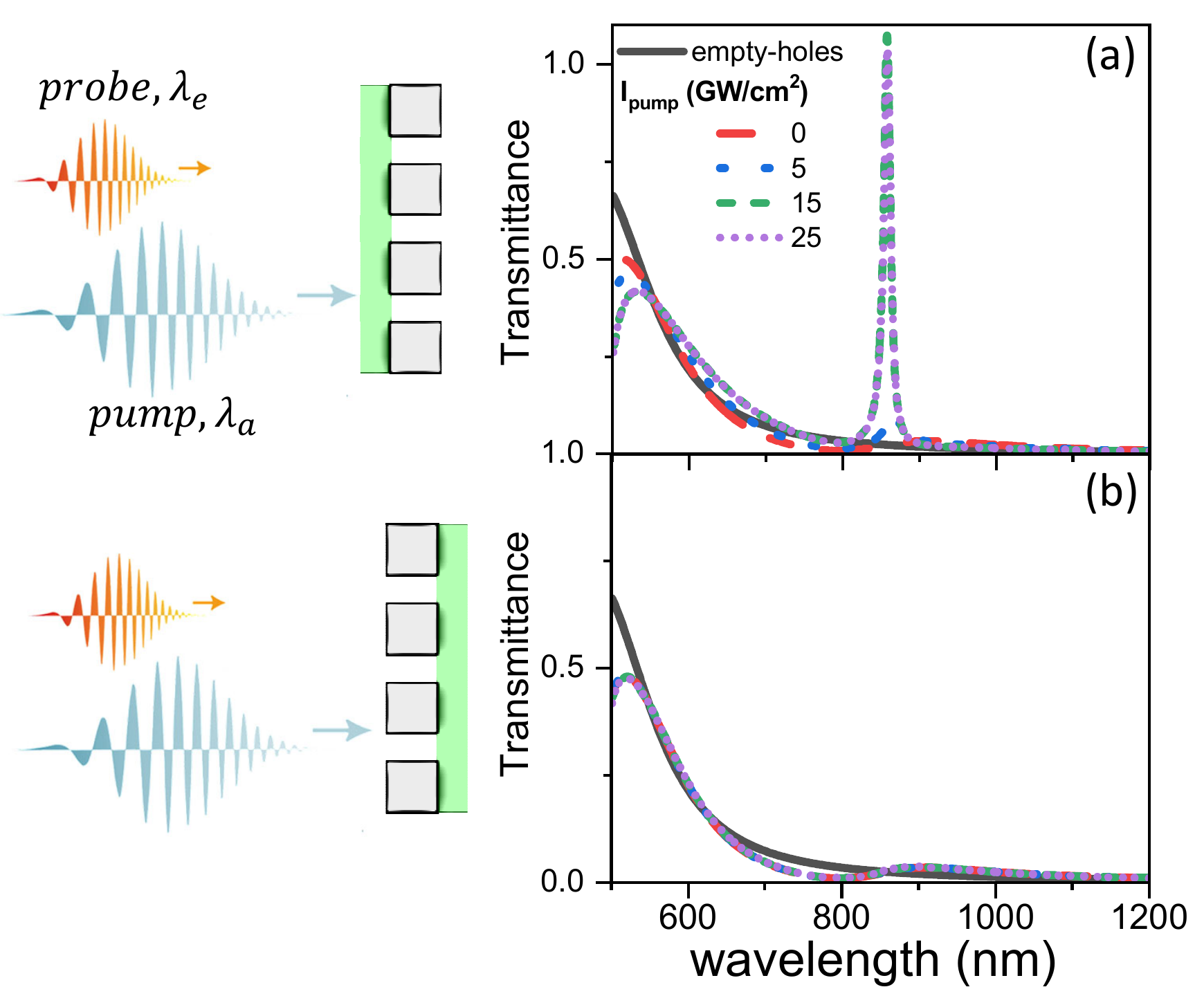} \caption{\textbf{Surface AIT:} Probe transmittance through an AIT metasurface for different pump intensities and two sample orientations. The system is pumped and probed from the same region where the IR-140 molecules are deposited in (a), while the thin slab of dye laser is on the other side in (b). The passive case (no dye in the structure) is shown with a solid black line in each case. The AIT metasurface is similar to the one of Fig.~\ref{fig1} but for another hole size, here $d = 220$~nm.} \label{fig2}
\end{figure}

AIT can also take place when the molecules are not located inside the holes but on the surface, because the excitation of SPPs~\cite{ZhongACSNano16}.  In this case there is an interesting result, the metasurface response resembles to the behavior that of an optical diode. Figure~\ref{fig2} shows the probe transmittance for different pump intensities (see labels) and for two sample orientations regarding the pump-probe direction (the geometrical parameters are similar to the ones of localized AIT, except the hole size that is $d = 220$~nm in this case). In Fig.~\ref{fig2}(a) peaks at the emission energy of IR-140 dye are observed when  pumping the array of nanoholes from the region where the dye laser is deposited. However this peak is absent when the pump illuminates the system from the other side, shown in Fig.~\ref{fig2}(b). From the point of view of the probe, the metasurface before pumping is opaque, no matter if it impinges the system from the side free of molecules or from the other side. However, after pumping the metasurface becomes transparent through the side covered by molecules, while transmission is still negligible and comparable to the passive case otherwise.

To understand the origin of the AIT peak and the relation with the propagation constant of the SPPs excited at the silver/dye-laser interface, we investigate the dispersion relation of a single silver/dye-laser interface (semi-infinite media). Of course the last is an approximation, but due to the extraordinary vertical confinement of the SPPs excited in the actual situation the semi-infinite approach suffices to illustrate the physics behind surface AIT. The dispersion relation of SPPs is:

\begin{equation}\label{spp}
	k_{spp}(\omega)=\frac{\omega}{c} \sqrt{\frac{\varepsilon(\omega) \varepsilon_m(\omega)}{\varepsilon(\omega)+\varepsilon_m(\omega)}}
\end{equation}
where $\varepsilon(\omega)$ and $\varepsilon_m(\omega)$ are the dielectric constants of the gain medium and the metal, respectively. 

\begin{figure}
	\centering\includegraphics[width=1.0\columnwidth]{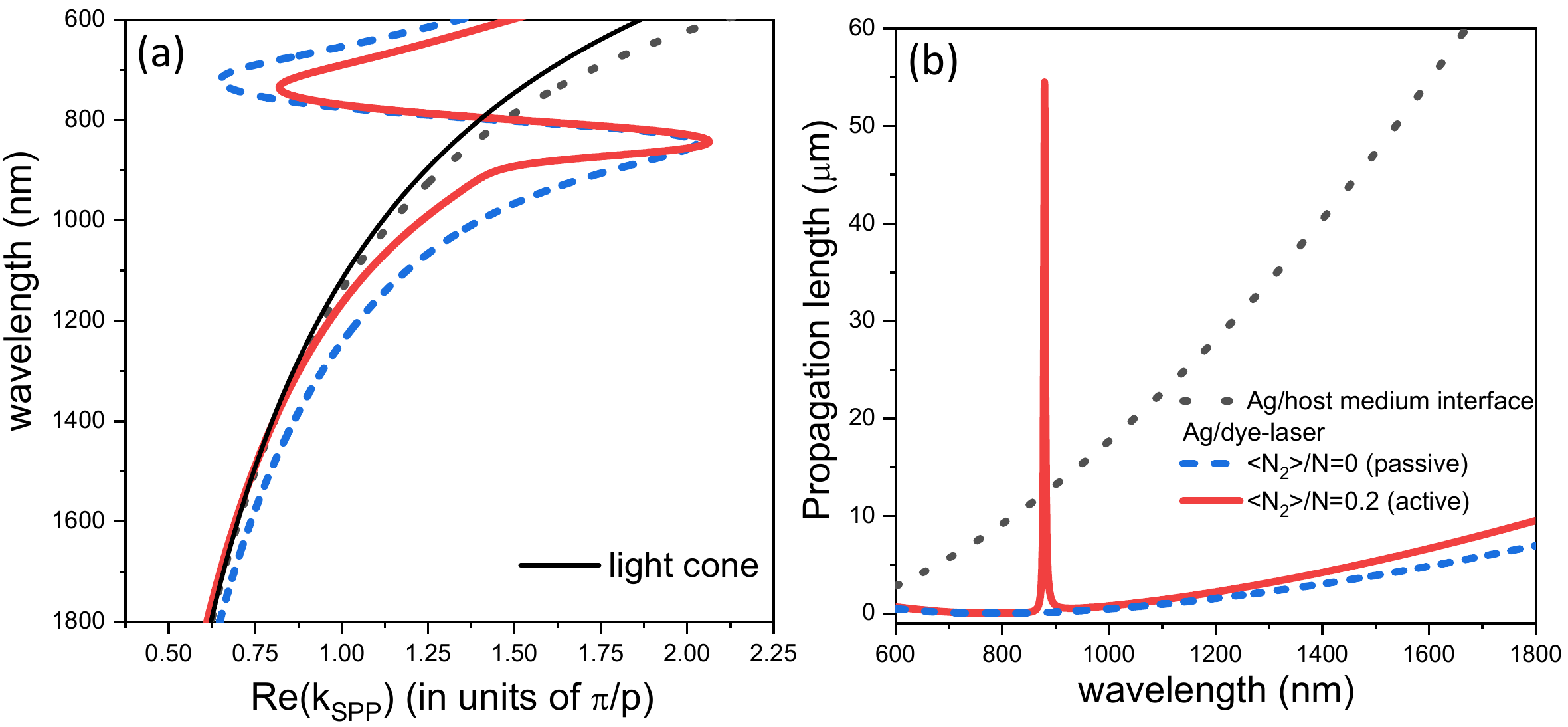} \caption{(a) The SPP dispersion relation at the silver/host-medium interface (gray line) is compared to the one for a silver/dye-laser interface, before (blue) and after (red) pumping with an intense laser. The laser populates the emission level up to 20\% of the available atomic states. The black line represents the light cone. (b) Corresponding propagation constant of SPPs on the different interfaces, as a function of the wavelength.} \label{fig3}
\end{figure} 

Fig.~\ref{fig3}(a) shows the dispersion relation, $Re(k_{spp})$, of a silver/dye-laser interface before (blue) and after (red) pumping the system with an intense laser. The corresponding silver/host-medium interface (before introducing the dye laser) is depicted with a gray line. In the example,  a population of the upper level of the emission transition of 20\% has been chosen to illustrate what it would occur in the actual structure. 

The real part of the propagation constant shows a "flat" region for the active case at the emission energy of the molecules (Fig.~\ref{fig3}(a)), which is related with a small group velocity of the SPP excited. The propagation length ($1/(2 Im(k_{spp})$), shown in Fig.~\ref{fig3}(b), changes from a few nanometers in the passive case to more than 50$\mu m$ in the active situation, again at $\lambda_e$. In the passive case damping is so high that the presence of these EM modes are difficult to observe in experiments with arrays of nanoholes. These SPP modes suffer of a so strong quenching that only with Surface Plasmon Resonance experiments have been described~\cite{YehAnal2012}. However the combination of a reduced group velocity and increased propagation length of the SPPs enhances the mechanism of EOT in the active case, which explain the results of Fig.~\ref{fig2}.

Finally, the paragraph above can help to explain something that has not been sufficiently considered in literature: a SPP resonance is always expected to occur at the $\Sigma$ point of the band structure in arrays of nanoles~\cite{ZhongACSNano16}. This phenomenon happens at the energy of the molecular resonance, due to the ''flatness'' of the dispersion relation at that energy (see Fig.~\ref{fig3}(a)). At the end, the most feasible way to investigate these EM modes is by exploiting the gain properties of the materials, like we have done in this work.

\section{Conclusions}\label{conclusions}
We have demonstrated how to actively control AIT metasurfaces through the modification of the optical response of an infrared dye laser deposited on top of them. An intense laser is able to modify the local dielectric constant of the gain medium, and in turn, the propagation of light either inside the holes and/or on the metal/gain-material interface. We have shown that optical diode-like behavior can be achieved for the probe field in surface AIT by externally acting on the propagation constant of SPPs. Because AIT can be found in other frequency bands, new avenues to active control in nanophotonics are possible in other spectral regimes, after this work. Finally, we believe that active AIT might be useful to build ultra-small lasers and other active photodevices for future nanothechnology. 

\vspace{0.2cm}

\begin{acknowledgments}
	This research was funded by the Spanish Ministry Science, Innovation and Universities grant number MAT2017-88358-C3-2-R (AEI/FEDER,UE).	
\end{acknowledgments}


%

\end{document}